\PassOptionsToPackage{unicode}{hyperref}
\PassOptionsToPackage{hyphens}{url}
\PassOptionsToPackage{dvipsnames,svgnames,x11names}{xcolor}
\documentclass[
]{article}
\usepackage{amsmath,amssymb}
\usepackage{lmodern}
\usepackage{iftex}
\ifPDFTeX
  \usepackage[T1]{fontenc}
  \usepackage[utf8]{inputenc}
  \usepackage{textcomp} 
\else 
  \usepackage{unicode-math}
  \defaultfontfeatures{Scale=MatchLowercase}
  \defaultfontfeatures[\rmfamily]{Ligatures=TeX,Scale=1}
\fi
\IfFileExists{upquote.sty}{\usepackage{upquote}}{}
\IfFileExists{microtype.sty}{
  \usepackage[]{microtype}
  \UseMicrotypeSet[protrusion]{basicmath} 
}{}
\makeatletter
\@ifundefined{KOMAClassName}{
  \IfFileExists{parskip.sty}{%
    \usepackage{parskip}
  }{
    \setlength{\parindent}{0pt}
    \setlength{\parskip}{6pt plus 2pt minus 1pt}}
}{
  \KOMAoptions{parskip=half}}
\makeatother
\usepackage{xcolor}
\providecommand{\tightlist}{%
  \setlength{\itemsep}{0pt}\setlength{\parskip}{0pt}}
\NewDocumentCommand\citeproctext{}{}
\NewDocumentCommand\citeproc{mm}{%
  \begingroup\def\citeproctext{#2}\cite{#1}\endgroup}
\makeatletter
 \let\@cite@ofmt\@firstofone
 \def\@biblabel#1{}
 \def\@cite#1#2{{#1\if@tempswa , #2\fi}}
\makeatother
\newlength{\cslhangindent}
\newlength{\csllabelwidth}
\newenvironment{CSLReferences}[2] 
 {\begin{list}{}{%
  \setlength{\itemindent}{0pt}
  \setlength{\leftmargin}{0pt}
  \setlength{\parsep}{0pt}
  \ifodd #1
   \setlength{\leftmargin}{\cslhangindent}
   \setlength{\itemindent}{-1\cslhangindent}
  \fi
  \setlength{\itemsep}{#2\baselineskip}}}
 {\end{list}}
\usepackage{calc}

\ifLuaTeX
\usepackage[bidi=basic]{babel}
\else
\usepackage[bidi=default]{babel}
\fi
\babelprovide[main,import]{american}

\def\languageshorthands#1{}
\ifLuaTeX
  \usepackage{selnolig}  
\fi
\IfFileExists{bookmark.sty}{\usepackage{bookmark}}{\usepackage{hyperref}}
\IfFileExists{xurl.sty}{\usepackage{xurl}}{} 
\hypersetup{
  pdftitle={pylhe: A Lightweight Python interface to Les Houches Event
files},
  pdfauthor={Alexander Puck Neuwirth, Matthew Feickert, Lukas Heinrich,
Eduardo Rodrigues},
  pdflang={en-US},
  colorlinks=true,
  linkcolor={Maroon},
  filecolor={Maroon},
  citecolor={Blue},
  urlcolor={Blue},
  pdfcreator={LaTeX via pandoc}}

\title{pylhe: A Lightweight Python interface to Les Houches Event files}

\definecolor{c53baa1}{RGB}{83,186,161}
\definecolor{c202826}{RGB}{32,40,38}

\usepackage[affil-it]{authblk}
\usepackage{orcidlink}
\author[1,2%
  \ensuremath\mathparagraph]{Alexander Puck Neuwirth%
    \,\orcidlink{0000-0002-2484-1328}\,%
    }
\author[3%
  ]{Matthew Feickert%
    \,\orcidlink{0000-0003-4124-7862}\,%
    }
\author[4%
  ]{Lukas Heinrich%
    \,\orcidlink{0000-0002-4048-7584}\,%
    }
\author[5%
  ]{Eduardo Rodrigues%
    \,\orcidlink{0000-0003-2846-7625}\,%
    }

\affil[1]{University of Milan Bicocca%
  }
\affil[2]{INFN Milan Bicocca%
  }
\affil[3]{University of Wisconsin-Madison%
  }
\affil[4]{Technical University of Munich%
  }
\affil[5]{University of Liverpool%
  }
\affil[$\mathparagraph$]{Corresponding author: %
}
\date{24 September 2026}

\begin{document}
\maketitle

\section{Summary}\label{summary}

\texttt{pylhe} is a lightweight Python library that provides a simple
and efficient interface for reading and writing Les Houches Event (LHE)
files, a standard format used by Monte Carlo event generators in
high-energy physics (\citeproc{ref-Alwall:2006yp}{Alwall \& others,
2007}). The library enables memory-efficient streaming of events from
XML format (\texttt{.lhe}, compressed \texttt{.lhe.gz}) and HDF5 format
(\texttt{.h5}, \texttt{.hdf5}, \texttt{lheh5}) files through a pythonic
iterator interface, allowing researchers to process arbitrarily large
event files without loading all events into memory simultaneously. It is
a standalone, event-generator-independent package, installable from PyPI
and Conda. \texttt{pylhe}, part of the Scikit-HEP community project
(\citeproc{ref-Rodrigues:2020syo}{Rodrigues \& others, 2020};
\citeproc{ref-Scikit-HEP_site}{\emph{{The Scikit-HEP Project}}, 2016}),
is well maintained and integrated within the broader scientific Python
ecosystem.

Historically, the first standards for event representation in
high-energy physics were the HEPEVT and HEPRUP common blocks
(\citeproc{ref-Boos:2001cv}{Boos \& others, 2001}), which provided a
Fortran-based structure for storing event information. As the complexity
of Monte Carlo event generators increased, the need for a more flexible
and extensible format led to the development of the LHE file format. LHE
introduced an XML structure that allowed for better organization of
event data and facilitated interoperability between different tools.
Typically, LHE files are used to describe parton-level events generated
by matrix element generators that are passed to parton shower and
hadronization programs producing HepMC files
(\citeproc{ref-Buckley:2019xhk}{Buckley et al., 2021};
\citeproc{ref-hans_dembinski_2022_7013498}{Dembinski et al., 2022};
\citeproc{ref-Dobbs:2001ck}{Dobbs \& Hansen, 2001}), or directly to
analysis frameworks such as Rivet
(\citeproc{ref-Bierlich:2024vqo}{Bierlich et al., 2024};
\citeproc{ref-Bierlich:2019rhm}{Bierlich \& others, 2020};
\citeproc{ref-Buckley:2010ar}{Buckley et al., 2013}). In contrast to LHE
files, HepMC stores fewer details about the particles, but also contains
many more particles per event. Both files are plain text-based formats,
making them human-readable but potentially large. Consequently,
compression using the \texttt{gzip} library is common practice for both
formats.

Since \texttt{pylhe} version 2.0.0, the library also supports the
HDF5-based (\citeproc{ref-hdf5}{The HDF Group, 1998}) LHEH5 format
(\citeproc{ref-Bothmann:2023ozs}{Bothmann et al., 2024};
\citeproc{ref-Hoche:2019flt}{Höche et al., 2019}), which allows for
significant improvements in write performance while keeping file sizes
for event storage under control.

The LHE format stores information within
\texttt{\textless{}init\textgreater{}} and multiple
\texttt{\textless{}event\textgreater{}} blocks consisting of
whitespace-separated values designed historically for straightforward
parsing in Fortran. Further details can be found in the original
definition of the Les Houches Event file standard
(\citeproc{ref-Alwall:2006yp}{Alwall \& others, 2007}). Following the
original publication there were two extensions to the LHE format,
version 2.0 in 2009 (\citeproc{ref-Butterworth:2010ym}{Butterworth \&
others, 2010}) and version 3.0 in 2013
(\citeproc{ref-Andersen:2014efa}{Andersen \& others, 2014}).
\texttt{pylhe} completely implements the widely adopted version 3.0,
that is the addition of \texttt{\textless{}scales\textgreater{}},
\texttt{\textless{}generator\textgreater{}}, and multiple weights via
\texttt{\textless{}initrwgt\textgreater{}},
\texttt{\textless{}rwgt\textgreater{}},
\texttt{\textless{}weight\textgreater{}},
\texttt{\textless{}weights\textgreater{}},
\texttt{\textless{}wgt\textgreater{}}, and
\texttt{\textless{}weightgroup\textgreater{}} on top of the version 1.0.
Features that were introduced in version 2.0 but removed in version 3.0,
such as \texttt{\textless{}pdfinfo\textgreater{}} or
\texttt{\textless{}clustering\textgreater{}}, are not supported.

\section{Statement of need}\label{statement-of-need}

The LHE format is used by all major Monte Carlo event generators such as
MadGraph (\citeproc{ref-Alwall:2014hca}{Alwall et al., 2014}),
POWHEG-BOX (\citeproc{ref-Alioli:2010xd}{Alioli et al., 2010};
\citeproc{ref-Frixione:2007vw}{Frixione et al., 2007};
\citeproc{ref-Nason:2004rx}{Nason, 2004}), Sherpa
(\citeproc{ref-Sherpa:2019gpd}{Bothmann \& others, 2019};
\citeproc{ref-Gleisberg:2008ta}{Gleisberg et al., 2009}), HERWIG
(\citeproc{ref-Bahr:2008pv}{Bahr \& others, 2008};
\citeproc{ref-Bellm:2015jjp}{Bellm \& others, 2016},
\citeproc{ref-Bellm:2019zci}{2020}; \citeproc{ref-Bewick:2023tfi}{Bewick
\& others, 2024}; \citeproc{ref-Corcella:2000bw}{Corcella et al.,
2001}), Pythia (\citeproc{ref-Bierlich:2022pfr}{Bierlich \& others,
2022}; \citeproc{ref-Sjostrand:2006za}{Sjostrand et al., 2006},
\citeproc{ref-Sjostrand:2007gs}{2008};
\citeproc{ref-Sjostrand:2014zea}{Sjöstrand et al., 2015}), Whizard
(\citeproc{ref-Kilian:2007gr}{Kilian et al., 2011};
\citeproc{ref-Moretti:2001zz}{Moretti et al., 2001}). While interfaces
for C/C++/Fortran exist in the respective generators, a lightweight,
standalone and easy-to-use Python interface equally supporting all these
event generators was missing until the inception of \texttt{pylhe} in
2015. Additionally, \texttt{pylhe} can serve as a crucial interface for
emerging machine learning applications in particle physics, allowing
researchers to efficiently extract event data for training neural
networks and other machine learning models used in event classification,
anomaly detection, and physics analysis. Past experience and the
ever-growing number of publications citing \texttt{pylhe} have shown
that the portfolio of users goes from experimentalists to
phenomenologists, some of which are themselves developers of event
generators.

As a library, \texttt{pylhe} is straighforwardly installable via the
standard means, and is maintained within the Scikit-HEP community
project that is very well integrated within the broader scientific
Python ecosystem.

\section{State of the field}\label{state-of-the-field}

Unlike the existing C/C++/Fortran interfaces provided by the Monte Carlo
event generators, \texttt{pylhe} offers a standalone, pure Python
interface that is more accessible and easier to use. When \texttt{pylhe}
was first developed, there were no other standalone Python libraries
available for reading and writing LHE files. However, a Python parser is
distributed as part of MadGraph (\citeproc{ref-Alwall:2014hca}{Alwall et
al., 2014}), while MadAnalysis 5 exposes its C++ LHE reader and writer
through a Python interface (\citeproc{ref-Conte:2012fm}{Conte et al.,
2013}). Nowadays, there are a few other smaller Python libraries with
less adoption than \texttt{pylhe}, which provide only read functionality
and are no longer actively maintained, such as \texttt{lhereader}
(\citeproc{ref-lhereader2}{Albert, 2021};
\citeproc{ref-lhereader1}{Biswas, 2017}). For completeness, it should be
mentioned that several LHE libraries exist in other programming
languages, such as Go (\texttt{go-hep}, \citeproc{ref-Binet2017}{Binet
et al., 2017}), Rust (\texttt{lhe}, \citeproc{ref-lhe-rs}{Fukuda, 2018};
\texttt{lhef}, \citeproc{ref-lhef}{Maier, 2024b};
\citeproc{ref-lhef-rs}{Weber, 2018}; \texttt{event\_file\_reader},
\citeproc{ref-event_file_reader}{Maier, 2024a}), Julia
(\texttt{LHEF.jl}, \citeproc{ref-LHEF_jl}{Fischer \& Ling, 2023}) and
Haskell (\texttt{lhe.hs}, \citeproc{ref-lhehs}{Lawrence, 2019}). These
provide varying degrees of completeness, but none support the LHEH5
format yet.

\section{Software design}\label{software-design}

\texttt{pylhe} allows for easy reading and writing of \texttt{.lhe} and
\texttt{.lhe.gz} files in Python, enabling seamless integration into
modern data analysis workflows in high-energy physics. The pythonic
event yielding approach allows for memory-efficient processing of
arbitrarily large LHE files by streaming events one at a time rather
than loading all of them at once into memory. Internally, \texttt{pylhe}
uses \texttt{xml.etree.ElementTree} to parse the XML structure, since
using the \texttt{lxml} library did not provide a significant speed up.
For LHEH5 files the library uses \texttt{h5py}
(\citeproc{ref-collette2013pythonhdf5}{Collette, 2013}) to read and
write the HDF5 format, via the same pythonic data structures, making it
easy to convert between the standard XML and newer HDF5 formats.

The library facilitates quick validation of event files through
programmatic access to event structure and particle properties, making
it straightforward to perform sanity checks on generated events. This
can be done for example via the integration with Scikit-HEP's Awkward
Array package \texttt{awkward}
(\citeproc{ref-Pivarski_Awkward_Array_2018}{Pivarski et al., 2018})
through the \texttt{to\_awkward()} function, which converts LHE events
into columnar data structures optimized for vectorized operations and
efficient analysis of large datasets.

The library endeavours to support all major Monte Carlo event generators
and relies on example XML and HDF5 format files to ensure test coverage.

\section{Research impact statement}\label{research-impact-statement}

\texttt{pylhe} is regularly used in various research projects and
publications within high-energy physics. Notably, it has been cited in
Higgs studies (\citeproc{ref-Brehmer:2019gmn}{Brehmer et al., 2019};
\citeproc{ref-Feuerstake:2024uxs}{Feuerstake et al., 2025};
\citeproc{ref-Stylianou:2023tgg}{Stylianou \& Weiglein, 2024}), in
Supersymmetry (SUSY), Beyond the Standard Model (BSM) and dark matter
searches (\citeproc{ref-Anisha:2023xmh}{Anisha et al., 2023};
\citeproc{ref-Beresford:2024dsc}{Beresford et al., 2024};
\citeproc{ref-Beresford:2018pbt}{Beresford \& Liu, 2019};
\citeproc{ref-Cheung:2024oxh}{Cheung et al., 2024};
\citeproc{ref-Kling:2020iar}{Kling, 2020};
\citeproc{ref-Zhou:2022jgj}{Zhou \& Liu, 2022},
\citeproc{ref-Zhou:2024fjf}{2025}), and in forward physics studies
(\citeproc{ref-Kelly:2021mcd}{Kelly et al., 2022};
\citeproc{ref-Kling:2022ykt}{Kling et al., 2023};
\citeproc{ref-Kling:2020mch}{Kling \& Trojanowski, 2020}). It is also
employed in methodological studies involving machine learning techniques
for event generation and analysis
(\citeproc{ref-Brehmer:2019xox}{Brehmer et al., 2020};
\citeproc{ref-Kofler:2024efb}{Kofler et al., 2025}).

\section{AI usage disclosure}\label{ai-usage-disclosure}

Generative AI tools have been used in the development of this software
and writing of the manuscript:

\begin{itemize}
\tightlist
\item
  Github's copilot has been used in reviewing pull requests.
\item
  VScode's copilot has been used as an advanced autocomplete.
\item
  ChatGPT has been used to identify the most pythonic solutions in case
  of ambiguity.
\end{itemize}

All the results generated by these tools have been reviewed by the
authors and are correct to the best of our knowledge.

\section{Author order}\label{author-order}

The authors agreed to list Alexander Puck Neuwirth first, with the
remaining authors listed alphabetically by family name.

\section{Acknowledgements}\label{acknowledgements}

We would additionally like to thank the contributors of \texttt{pylhe}
and the Scikit-HEP community for their support.

\section*{References}\label{references}
\addcontentsline{toc}{section}{References}

\protect\phantomsection\label{refs}

\end{document}